\documentclass[pra,aps,amssymb,twocolumn,showpacs]{revtex4}
\usepackage[dvips]{graphicx}

\newcommand{\size}{7.5cm}

\begin{document}

\title{Decoherence from Internal Dynamics in Macroscopic Matter-Wave Interferometry}

\author{Sang Wook Kim$^1$ and Hwa-Kyun Park$^2$}

\affiliation{
$^1$Department of Physics Education and Research Center for Dielectric and Advanced Matter Physics \\
Pusan National University, Busan 609-735, Korea \\
$^2$Korea Institute for Advanced Study, 207-43 Cheongryangri-dong, Dongdaemun-gu, Seoul 130-722, Korea}

\date{\today}

\begin{abstract}

Usually, decoherence is generated from the coupling with an outer environment. However, a macroscopic 
object generically possesses its own environment in itself, namely the complicated dynamics of internal 
degrees of freedom. We address a question: when and how the internal dynamics decohere interference of 
the center of mass motion of a macroscopic object. We will show that weak localization of a macroscopic 
object in disordered potentials can be destroyed by such decoherence.

\end{abstract}


\maketitle


\section{Introduction}

Superposition of states lies at the heart of quantum mechanics and gives rise to many of
its paradoxes. Not only can a particle go through two paths simultaneously, but the 
wavefunction of a pair of particles flying apart from each other is also entangled into
a non-separable superposition of states. However, such strange phenomena have never been
observed in our macroscopic world. It has been an important question why and how quantum 
weirdness disappears in large systems \cite{Zurek91}.

Environment, usually described by a huge number of variables, can destroy coherence among
the states of a quantum system. This is decoherence. The environment is {\em watching} the path 
followed by the system, and thus suppressing interference effects and quantum weirdness. In 
macroscopic systems, such process is so efficient that we see only its final result: the classical 
world around us \cite{Joos00,Zurek03}. For truly macroscopic superpositions, decoherence occurs 
on a very short time-scale that it is almost impossible to observe quantum coherences. However, 
mesoscopic systems present the possibility of investigating the process of decoherence and the 
transition from quantum to classical behavior \cite{Haroche98}. So far many experiments have been 
realized to generate mesoscopic superpositions \cite{Brune96,Monroe96} and to decohere them in 
a controlled way \cite{Myatt00}. Recently considerably large molecules have been used to investigate 
the decoherence, the transition from quantum to classical. For example, the researchers in Wien
have observed interference of de Broglie waves of fullerenes (C$_{60}$ or C$_{70}$ molecules) and 
even bigger ones \cite{Arndt99,Nariz01,Brezger02,Hornberger03,Hackermueller03,Hackermueller04}. 
In this experiment the fullerenes are quite hot as well as big, which means they contain complicated 
dynamics of their internal degrees of freedom. The internal thermal energy is almost one order 
of magnitude larger than the kinetic energy of its center of mass (CM) motion. A question naturally 
arises: is the complex dynamics of the internal degrees of freedom harmful for the interference of 
the CM motion?

Usually, decoherence has been generated from the coupling with an {\em outer} environment such 
as other particles or fluctuating electromagnetic fields. However, a macroscopic object generically 
possesses its own environment in itself, namely the {\em internal} dynamics (ID) \cite{Kolovsky94a} 
when only small part of the total system, e.g. its CM, is under consideration. In this paper, 
we would like to address a question: when and how the ID decoheres interference of a macroscopic 
object. We also show nontrivial expectation that the weak localization of large molecules in a disordered 
potential can be destroyed by the decoherence generated from the ID without any external perturbation 
breaking the time reversal symmetry.

Let us consider a macroscopic object consisting of $N$ particles exposed to the external potential $V_{ex}$. 
The total Hamiltonian can be written as
\begin{eqnarray}
H & = & \sum_i^N \frac{p_i^2}{2m_i} + V_{in}(\{x_i\}) + \sum_i^N V_{ex}(x_i) \nonumber \\
  & = & \frac{P^2}{2M} + V_{ex}(X) + \sum_\alpha^{N-1} \frac{\mu_\alpha^2}{2m_\alpha} 
+ V_{in}(\{\xi_\alpha\}) + \Gamma\left(X,\{\xi_\alpha\}\right) \nonumber \\
  & \equiv & H_{CM} + H_{in} + \Gamma,
\end{eqnarray}
where $V_{in}$ is an internal (or confinement) potential. $P$ and $X$ are a momentum and a coordinate of 
the CM with a mass $M (=\sum m_i)$ respectively, while $\mu_\alpha$ and $\xi_\alpha$ the same 
quantities of the internal degrees of freedom with the reduced mass $m_\alpha$. $\Gamma$ represents 
the coupling between the CM and the ID. Since one finds $V_{in}$ is determined only from the relative 
coordinates $\xi_\alpha$, i.e. $V_{in}(\{x_i\}) = V_{in}(\{\xi_\alpha\})$, the coupling term 
$\Gamma$ depends only on the external potential $V_{ex}$. It is easy to show that if $V_{ex}$ does 
not correspond to a simple form such as constant, linear, and harmonic, the non-zero coupling 
$\Gamma$ always exists \cite{Park03}. {\em The CM motion can be entangled with its ID when 
the anharmonic external potential is applied.} We call such a non-trivial external potential 
``nonlinear'' since the corresponding force is nonlinear. It is noted that existence of the external 
potential has nothing to do with generation of the decoherence.

In a usual two slit experiment, a macroscopic object freely flies to a screen. Therefore, no entanglement 
between CM and ID arises, neither does the decoherence from ID. However, one can ask what happens 
if the repulsive potential produced by the slits is considered. For example, the van der Waals interaction 
between the C$_{70}$ and the grating was assumed to correctly explain experimental results \cite{Brezger02}. 
Since C$_{70}$ molecule is too small to see the anharmonic shape of the external potential the coupling 
between the CM and the ID is hardly expected. One can still insist, however, that not only if the slit wall 
contains more rapidly varying repulsive potential but also if the size of the object becomes larger, for example 
by using an insulin, then the coupling term might manifest itself. In this paper we consider such situation that 
the coupling between CM and ID is non-negligible. It must be mentioned that we ignore all the other sources 
of decoherences except those originating from the ID for our discussion, but will briefly comment it. We also 
note that neither we derive any new formula nor estimate any values of this effect. We would like to 
show basic physical mechanism and possibility of the phenomena.

\begin{figure}
\center
\includegraphics[height=\size,angle=0]{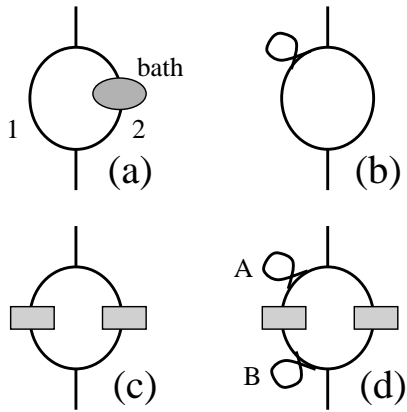}
\caption{Various setups of two-slit interferometry (a) with the path 2 containing a usual external bath
denoted by the grey circular region, (b) with no bath but delay, (c) with two paths containing the same 
nonlinear regions denoted by the grey boxes. (d) The same as (c) with delay in one of the paths after 
(A) and before (B) the interaction within the nonlinear region.}
\label{fig1}
\end{figure}

\section{Review of decoherence}

First, we briefly review the decoherence from an external environment. Figure 1(a) shows usual two path 
interferometry, where one of the paths interacts with a ``bath'' of particles, i.e. an environment.
One can write an initial  total wavefunction in the following way
\begin{equation}
\Psi (0) = \frac{1}{\sqrt{2}}\left[\psi_1 (0,X) +\psi_2 (0,X) \right]\otimes \chi (0,\xi),
\end{equation}
where $\psi_{1(2)}$ and $\chi$ denote a wavefunction of a system moving along the path 1(2) and that of a 
bath described by variables $\xi$, respectively. Under the dynamics, which includes interaction with the bath 
only along the path 2, the wavefunction becomes
\begin{equation}
\Psi (t) = \frac{1}{\sqrt{2}}\left[\psi_1 (t,X) \otimes \chi_1 (t,\xi) +\psi_2 (t,X) \otimes \chi_2 (t,\xi)\right],
\label{final_state}
\end{equation}
where $\chi_{1(2)}$ is a state of the bath with the system going around the path 1(2). Here, we assume 
the coupling with the bath is small enough to have no influence on the system, but changes only a state of 
the bath \cite{Stern90,Fiete03}. 

From the reduced density matrix the interference term is given as
\begin{equation}
{\rm Re}\left[ \psi_1^* (t,X) \psi_2 (t,X) \int d\xi \chi_1^* (t,\xi) \chi_2 (t,\xi) \right].
\label{interference}
\end{equation}
The physical meaning of Eq.~(\ref{interference}) is obvious. The first term contains usual information 
of the interference pattern of the system going through the two paths. The second term corresponds to 
the visibility which measures the decoherence. Eq.~(\ref{interference}) allows one to interpret the 
reduction of the interference in terms of a reduction in the overlap of the bath states for the two paths. 
Stern, Aharony and Imry \cite{Stern90} argue that one can make the identification
\begin{equation}
\left< e^{i\hat{\phi}} \right> = \int d\xi \chi_1^* (t,\xi) \chi_2 (t,\xi),
\label{decoh_phase}
\end{equation}
where $\left< e^{i\hat{\phi}} \right> \equiv \int d\xi \chi^* (0,\xi) e^{i\phi(\xi)}\chi (0,\xi)$, and
$\phi(\xi) = -\int dt V_I(t,X_2(t),\xi)$. Here $V_I$ denotes the interaction potential between the system 
and the environment in the interaction picture, i.e. $V_I \equiv \exp(iH_{env}t)V(X_2,\xi)\exp(-iH_{env}t)$
in which $H_{env}$ and $V$ represent the Hamiltonian of the environment and its interaction with the system,
respectively. Eq.~(\ref{decoh_phase}) implies that the reduction of the interference can also be ascribed 
to the accumulating phase uncertainty of the system on the interacting path being subject to an uncertain 
potential. In this sense decoherence is often referred to as ``dephasing''. 

\section{Decoherence from internal dynamics}

\subsection{Two slit interferometry}

Now let us consider two path interferometry of a macroscopic object with its complicated ID. The CM and the ID 
now play roles of a system and an environment, respectively. We take into account the case that the object is 
moving freely as shown in Fig.~1(b), so that there is no entanglement between the CM and the ID. The motion
of the CM is easily described by a plane wave, $\psi_0(t,X)$. The CM of the object then shows perfect coherence 
since the CM dynamics is completely isolated. When one increases the path length of one of the two paths
by amount of $l$, clear interference pattern is expected from the term 
$\psi_1^* (t,X) \psi_2 (t,X) = \psi_0^* (t+\Delta t,X) \psi_0 (t,X)$ 
between two CM states, where $\Delta t =l/v$ ($v$ is the velocity of the CM). The final wavefunction is 
given as
\begin{equation}
\Psi (t) = \frac{1}{\sqrt{2}}\left[\psi_0 (t+\Delta t,X) + \psi_0 (t,X) \right] \otimes \chi (t,\xi).
\label{final_state_no}
\end{equation}
The time delay $\Delta t$ is only applicable to the CM since the ID independently evolves in time. 
The overlap of the ID always yields that of the same states, i.e. complete coherence. Without 
entanglement with the ID the decoherence of the CM cannot be generated. 

Let us consider now the case an object goes though {\em nonlinear} external potentials along the paths as shown 
in Fig.~1(c). We assume these nonlinear potentials are equivalent for the two paths. In a two slit interferometry, 
in general, the two slits are made to have the same geometry. Even though the CM motion is entangled with the 
ID during the passage through such nonlinear region, this does not generate any decoherence. The reason is 
that the evolution of the ID is always equal for the two paths, so that $\chi_1(t,\xi) = \chi_2 (t,\xi)$,
i.e. $\int d\xi \chi_1^* (t,\xi) \chi_2 (t,\xi) = 1$. The entanglement with internal environment
makes the phases of each of the partial waves of the CM uncertain in viewpoint of Eq.~(\ref{decoh_phase}), 
but does not alter the relative phase.  In Fig.~1(d), we introduce additional delay for one of the paths. 
When the delay is given after the passage through the nonlinear region, the situation exactly corresponds 
to the case described in Fig.~1(b). The only difference is that one starts not with an initial ID state 
$\chi_1 (0,\xi)$ [$=\chi_2 (0,\xi)$] but with $\chi_1 (t_0,\xi)$ [$=\chi_2 (t_0,\xi)$], where $t_0$ is the 
time when the object departures from the nonlinear region. When the delay is given before, one can see that 
the situation is also the same as the case shown in Fig.~1(b) by considering the argument related to 
Eq.~(\ref{final_state_no}). 

It should be noted that in usual decoherence generated from an outer environment it is not easy to find the case that 
the two paths have the same environment. If the decoherence occurs mainly by interacting with other particles, the 
system going through two paths accumulates different random phases from scattering with different particles of different 
states. The system going through the two paths thus see different environments. This is the reason why the decoherence
from an outer environment has been dealt with the setup shown in Fig.~1(a). One example that the same environment is
applied to the two interfering waves is the interaction of an interfering electron with zero-point (or vacuum) 
fluctuation, where the electron does not decohere \cite{Kumar87,Rammer88,Imry02}.

\subsection{Quantification of decoherence from internal dynamics}

In the above discussion it has been shown that it is not easy to see the decoherence generating from the ID with usual 
simple geometry of interferometry. The only way to observe the decoherence from the coupling with ID is that the ID's 
should see different nonlinear interactions for each path. Without loss of generality this situation can be represented as 
the case that only one path contains the nonlinear region. The overlap of the ID can then be written as
\begin{equation}
\left< \chi (0,\xi) \right| e^{iH_{in}\Delta t} 
{\it \hat{T}} \exp\left[-i \int_0^{\Delta t} dt\left( H_{in}+\Gamma\right)\right] \left| \chi (0,\xi) \right>,
\label{fidelity}
\end{equation}
where $\Delta t$ and ${\it \hat{T}}$ denote the interaction time within the nonlinear potential and the time ordering 
operator, respectively. Asymmetric geometry of interferometry is sometimes useful, for example, measurement of phase 
of the transmission coefficient through a quantum dot \cite{Schuster97}, where a quantum dot is plugged into one of 
the arms of an Aharonov-Bohm ring. From the Aharonov-Bohm oscillation one can determine the phase shift of electron
passing through the quantum dot.

The quantity given in Eq.~(\ref{fidelity}) is known as so called  {\em fidelity} \cite{Nielsen00}. The decay of such a 
quantity determines the decoherence rate. One important remark is that even for the ID with a few degrees of freedom the 
fidelity decays exponentially if its dynamics is chaotic \cite{Peres84,Jalabert01}. It opens possibility that the decoherence 
can occur in molecules consisting of even several atoms from entanglement with its ID \cite{Park03} (See also 
\cite{Adachi88,Kolovsky94b,Kubotani95,Nakazato96,Tanaka02,Cohen04} for decoherence generated from chaos) in a 
certain special condition. When a single coherent state, a minimal wavepacket, is chosen as an initial state of ID, one can 
expect golden rule or Lyapunov decay depending on the strength of the coupling \cite{Jacquod01}, in which the fidelity 
decay does not much depend on the initial condition for a given energy as far as the completely chaotic dynamics is 
concerned for the ID. This situation corresponds to the ID governed by rather small number of degrees of freedom at low 
temperature. As we mentioned in the beginning, however, for small molecules it is hard to expect the coupling between
the CM and the ID. To see this effect in the systems with small degrees of freedom we need something different from
usual molecules.

For rather bigger systems, which we are interested in, it is not easy to directly calculate how much the interaction between
the CM and the ID make an influence on the state of the ID because the internal degrees of freedom consist of many particles. 
First let us consider the zero temperature case. If the interaction is strong enough to generate any kind of elementary excitations 
such as phonons, charge density waves, magnons for magnetic systems, and so on, then the CM will lose his coherence 
completely. Nothing happens for the system going through the path 1 in Fig.~\ref{fig1}, while the state of the ID through the 
path 2 in Fig.~\ref{fig1} is changed from the ground state to the excited state of the elementary excitation. By checking the state 
of the ID one can see which path the system go through. It is nothing but a complete decoherence. In this case it is crucial to 
know whether the excitation is gapless or 
not. At finite temperature the situation is much more complicated. In the beginning we assumed that there is no other sources 
of decoherences except those originating from the ID. Unfortunately this is no longer true since an object with finite temperature
is always coupled with outer environment by emitting black body radiation or possibly other radiation from thermal vibration. 
It is another issue how much this effect degrades
the coherence of the motion of the CM. Now one needs to compare the decoherence from the coupling with the outer environment
and that with the ID. If the radiation is not so harmful for the coherence, e.g. the case that the wavelength of the radiation is much 
larger than the difference of two paths, it is meanigful to calculate Eq.~(\ref{fidelity}). This calculation can be done by using the 
field theoretical technique once the time dependent interaction $\Gamma$ is known. Surely it is still hard task. We do not want to 
calculate it in this paper, but only point out the possible existence of the decoherence generating from the coupling with the ID.

Another remark on Eq.~(\ref{fidelity}) is its physical interpretation in terms of quantum measurement. Here, 
the total system including CM and ID is isolated from the external world. The decoherence is generated from entanglement 
only within a macroscopic object itself. In this sense no external observer exists. No information is transferred 
from the object to outside.  Since a system (the CM) does not know the difference between an internal and an outer 
environment, the ID plays a role of an observer. Information thus flows from the CM to the ID. Following Zurek's 
argument \cite{Zurek03}, one can say {\em the ID is watching the CM}. 

\begin{figure}
\center
\includegraphics[height=\size,angle=0]{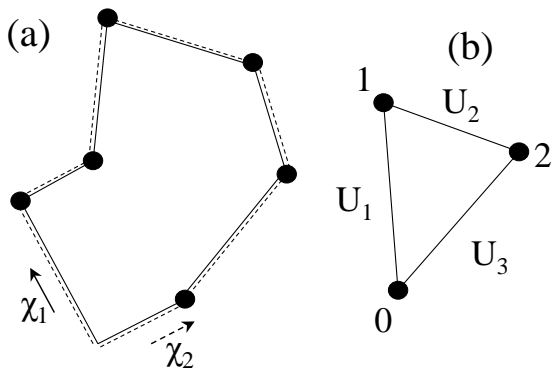}
\caption{(a) Two time reversal paths, $\chi_1$ and $\chi_2$, multiply reflecting from random scatterers in a 
disordered system. (b) Simple example with three scatterers labeled by 0, 1, and 2. $U_n$ denotes an unitary operator 
freely propagating between two scatterers.}
\label{fig2}
\end{figure}

\subsection{Weak localization in disordered systems}

Finally, let us consider one interesting example, weak localization in disordered systems \cite{Bergman84}. Two time reversal 
paths multiply reflecting from the random scatterers leads to the localization of a wavefunction due to their constructive 
interference as shown in Fig.~2(a). Such localization is fragile for both the decoherence and the perturbation breaking 
time reversal symmetry. A random potential generates complicated and chaotic dynamics, which can give rise to the 
entanglement between CM and ID. At first glance the decoherence from the ID is hardly expected to arise since {\em the total 
system has time reversal symmetry}. One might think that the state of the ID of the CM rotating clockwise must be the same 
as that of counter-clockwise. It can be shown, however, that the coupling to the ID can destroy the weak localization by 
generating the decoherence from the ID. 

To prove the appearance of the decoherence let us consider the overlap between two ID, namely $\chi_1$ and $\chi_2$, 
of the time reversal paths of CM after a round trip along a closed loop. Since we are interested only 
in the ID, the influence from its coupling to the CM can be regarded as {\em a time dependent external perturbation}, i.e. 
$\Gamma(X_{1(2)}(t),\xi) \equiv \Gamma_{1(2)}(t,\xi)$. Note $\Gamma_{1(2)}(t+\tau,\xi)=\Gamma_{1(2)}(t,\xi)$, 
where $\tau$ denotes the duration time taken for a round trip around the closed loop. The final state $\chi_{1(2)}(\tau)$ 
is then given as 
\begin{equation}
\chi_{1(2)} (\tau,\xi) = {\it \hat{T}}\exp\left\{-i \int_0^\tau dt\left[ H_{in}(\xi)+\Gamma(X_{1(2)}(t),\xi)\right]\right\} \chi_0
.\end{equation}
Even though $X_1(t)=X_2(\tau-t)$ holds, i.e. $\Gamma_1(t,\xi)=\Gamma_2(\tau-t,\xi)$, one finds $\chi_1(\tau) \neq \chi_2(\tau)$ 
due to existence of the time ordering operator ${\it \hat{T}}$. To make it more clear, let us consider a simple example: three scatterers well 
localized in space as shown in Fig.~2(b). During free propagation between two scatterers the CM is decoupled from the ID. We assume
the process of collision with the scatterers is short enough to be described by delta-function in time. The interaction term $\Gamma_1$ for 
the clockwise propagation can then be given as
\begin{equation}
\Gamma_1(t,\xi) = f_1(\xi)\delta(t-\tau_1) + f_2(\xi)\delta(t-\tau_2),
\end{equation}
where $\tau_1$ and $\tau_2$ are the collision times upon the first and the second scatterer, respectively, and 
$0 < \tau_1 < \tau_2 < \tau$. After one round trip, the states of the ID for the clockwise and the counter-clockwise 
$\chi_1$ and $\chi_2$ are respectively given as 
\begin{eqnarray}
\chi_1(\tau) = U_3 K_2 U_2 K_1 U_1 \chi_0 \nonumber \\
\chi_2(\tau) = U_1 K_1 U_2 K_2 U_3 \chi_0,
\end{eqnarray}
where by using the eigenstates of $H_0$ ($H_0 \left| i\right> = E_i \left| i\right>$) one obtains
$U_{n,ij} = \exp[-iE_i \Delta t_n] \delta_{ij}$, and $K_{n,ij} = \left< i|\exp[-if_n(\xi)]|j\right>$.
Here, $\Delta t_1 = \tau_1$, $\Delta t_2 = \tau_2 - \tau_1$, and $\Delta t_3 = \tau - \tau_2$. It is obvious that
in general $ U_3 K_2 U_2 K_1 U_1 \neq U_1 K_1 U_2 K_2 U_3$ since $K_1$ and $K_2$ are not diagonal. Consequently
one can find that in general $\left< \chi_1(\tau) | \chi_2(\tau) \right> < 1$. {\em The weak localization of the CM 
of a macroscopic object in disordered potentials can be destroyed due to the coupling to the ID}.

\section{Summary}

In summary, we have investigated the decoherence generated from the internal dynamics of a macroscopic object. In a 
usual setup of two slit interferometry, it is hard to expect the appearance of such decoherence. Only asymmetric geometry 
of the interfering paths containing anharmonic external potential allows one to observe the decoherence from the internal 
dynamics. Such decoherence can then be measured by the fidelity given in Eq.~(\ref{fidelity}). In this case, the internal 
degrees of freedom of a macroscopic object are watching its center of mass motion. The weak localization of the center of 
mass motion of a macroscopic object in disordered potentials can also be destroyed by such decoherence without any 
external perturbation breaking time reversal symmetry.

\section*{Acknowledgments}
SWK would like to thank Fritz Haake and Klaus Hornberger for helpful discussions. The part of this work has been done
when SWK has participated in the focus program of Asia Pacific Center for Theoretical Physics (APCTP) in Pohang, Korea.


\bibliographystyle{prsty}

\begin{thebibliography}{10}

\bibitem{Zurek91}
W.H. Zurek, Phys. Today {\bf 44} (10), 36 (1991).

\bibitem{Joos00}
E. Joos, in {\it Decoherence: Theoretical, Experimental and Conceptual Problems}
edited by P. Blanchard, D. Giulini, E. Joos, and I.-O. Stamatescu, Lecture Notes in Physics
Vol. 538 (Springer-Verlag, Heidelberg, 2000) p.1.

\bibitem{Zurek03}
W.H. Zurek, Rev. Mod. Phys. {\bf 75}, 715 (2003).

\bibitem{Haroche98}
S. Haroche, Phys. Today {\bf 51} (7), (1998).

\bibitem{Brune96}
M. Brune, E. Hagley, J. Dreyer, X. Maitre, A. Maali, C. Wunderlich, J.M. Raimond,
and S. Haroche, Phys. Rev. Lett. {\bf 77}, 4887 (1996).

\bibitem{Monroe96}
C. Monroe, D.M. Meekhof, B.E. King, and D.J. Wineland, Science {\bf 272}, 1131 (1996).

\bibitem{Myatt00}
C.J. Myatt, B.E. King, Q.A. Turchette, C.A. Sackett, D. Kielpinski, W.M. Itano,
and D.J. Wineland, Nature {\bf 403}, 269 (2000).

\bibitem{Arndt99}
M. Arndt, O. Nariz, J. Vos-Andrae, C. Keller, G. van der Zouw, and A. Zellinger,
Nature {\bf 401}, 680 (1999).

\bibitem{Nariz01}
O. Nariz, B. Brezger, M. Arndt, and A. Zellinger, Phys. Rev. Lett. {\bf 87}, 160401
(2001).

\bibitem{Brezger02}
B. Brezger, L. Hackerm\"uller, S. Uttenthaler, J. Petschinka, M. Arndt, and A. Zellinger,
Phys. Rev. Lett. {\bf 88}, 100404 (2002).

\bibitem{Hornberger03}
K. Hornberger, S. Uttenthaler, B. Brezger, L. Hackerm\"uller, M. Arndt, and A. Zellinger,
Phys. Rev. Lett. {\bf 90}, 160401 (2003).

\bibitem{Hackermueller03}
L. Hackerm\"uller, S. Uttenthaler, K. Hornberger, E. Reiger, B. Brezger, A. Zellinger, and M. Arndt,
Phys. Rev. Lett. {\bf 91}, 090408 (2003).

\bibitem{Hackermueller04}
L. Hackerm\"uller, K. Hornberger, B. Brezger, A. Zellinger, and M. Arndt,
Nature {\bf 427}, 711 (2004).

\bibitem{Kolovsky94a}
A.R. Kolovsky, Europhys. Lett. {\bf 27}, 79 (1994).

\bibitem{Park03}
H.-K. Park and S.W. Kim, Phys. Rev. A {\bf 67}, 060102(R) (2003).

\bibitem{Grisenti99}
R.E. Grisenti, W. Sch\"ollkopf, J.P. Toennies, G.C. Hegerfeldt, and T. K\"ohler,
Phys. Rev. Lett. {\bf 83}, 1755 (1999).

\bibitem{Stern90}
A. Stern, Y. Aharony, and Y. Imry, Phys. Rev. A {\bf 41}, 3436 (1990).

\bibitem{Fiete03}
G.A. Fiete and E.J. Heller, Phys. Rev. A {\bf 68}, 022112 (2003) .

\bibitem{Kumar87}
N. Kumar, D.V. Baxter, R. Richter, and J.O. Stomolsen, Phys. Rev. Lett. {\bf 59}, 1853 (1987).

\bibitem{Rammer88}
J. Rammer, A.L. Shelankov, and A. Schmid, Phys. Rev. Lett. {\bf 60}, 1985 (1988);
G. Bergman, Phys. Rev. Lett. {\bf 60}, 1986 (1988).

\bibitem{Imry02}
Y. Imry, cond-mat/0202044 (unpublished).

\bibitem{Schuster97}
R. Schuster, E. Buks, M. Heiblum,D. Mahalu, V. Umansky, and H. Shtrikman,
Nature {\bf 385}, 417 (1997).

\bibitem{Nielsen00}
H.M. Nielsen and I.L. Chuang, {\it Quantum Computation and Quantum Information}
(Cambridge University Press, Cambridge, 2000).

\bibitem{Peres84}
A. Peres, Phys. Rev. A {\bf 30}, 1610 (1984).

\bibitem{Jalabert01}
R.A. Jalabert and H.M. Pastawski, Phys. Rev. Lett. {\bf 86}, 2490 (2001).

\bibitem{Adachi88}
S. Adachi, M. Toda, and K. Ikeda, Phys. Rev. Lett. {\bf 61}, 659 (1988).

\bibitem{Kolovsky94b}
A.R. Kolovsky, Phys. Rev. E {\bf 50}, 3569 (1994).

\bibitem{Kubotani95}
H. Kubotani, T. Okamura, M. Sakagami, Physica A {\bf 214}, 560 (1995).

\bibitem{Nakazato96}
H. Nakazato, M. Namiki, S. Pascazio, and Y. Yamanaka, Phys. Lett. A {\bf 222}, 130 (1996).

\bibitem{Tanaka02}
A. Tanaka, H. Fujisaki, and T. Miyadera, Phys. Rev. E {\bf 66}, 045201(R) (2002).

\bibitem{Cohen04}
D. Cohen and T. Kottos, Phys. Rev. E {\bf 69}, 055201(R) (2004).

\bibitem{Jacquod01}
Ph. Jacquod, P.G. Silvestrov, and C.W.J. Beenakker, Phys. Rev. E {\bf 64}, 055203(R) (2001).

\bibitem{Bergman84}
G. Bergman, Phys. Rep. {\bf 107}, 1 (1984).

\end{thebibliography}

\end{document}